\documentclass[amsmath,amssymb,pra,aps,showpacs,superscriptaddress,twocolumn]{revtex4-2}
\usepackage{amsmath,amsfonts,amssymb,amsthm,graphics,graphicx,epsfig,bbm}
\usepackage[colorlinks=true,citecolor=blue,linkcolor=blue,urlcolor=blue]{hyperref}
\usepackage[usenames]{color}
\usepackage{graphicx}
\usepackage{subfigure}
\usepackage{amsmath}
\usepackage{epsfig}
\usepackage{dcolumn}
\usepackage{bm}
\usepackage{color}
\usepackage{mathdots}
\usepackage{times}
\usepackage{epstopdf}
\usepackage{amssymb}
\usepackage{amstext}
\usepackage{latexsym}
\usepackage{hyperref}
\usepackage{amsfonts}
\usepackage{psfrag}
\usepackage{soul,xcolor}
\usepackage[normalem]{ulem}
\usepackage{dsfont}
\usepackage{txfonts}
\usepackage{float}

\newcommand{\ket}[1]{\vert #1 \rangle}

\newcommand{\ketbra}[2]{\vert #1 \rangle \langle #2 \vert}
\newcommand{\braket}[2]{\langle #1 \vert #2 \rangle}

\usepackage{tikz,xcolor,hyperref}
\definecolor{lime}{HTML}{A6CE39}
\DeclareRobustCommand{\orcidicon}{%
	\begin{tikzpicture}
	\draw[lime, fill=lime] (0,0) 
	circle [radius=0.16] 
	node[white] {{\fontfamily{qag}\selectfont \tiny ID}};
	\draw[white, fill=white] (-0.0625,0.095) 
	circle [radius=0.007];
	\end{tikzpicture}
	\hspace{-2mm}
}

\foreach \x in {A, ..., Z}{%
	\expandafter\xdef\csname orcid\x\endcsname{\noexpand\href{https://orcid.org/\csname orcidauthor\x\endcsname}{\noexpand\orcidicon}}
}

\begin{document}

\setstcolor{red}

\title{Equally entangled multiqubit states} 
\date{\today}

\author{Francisco Albarr\'an-Arriagada \orcidA{}} 
\affiliation{Departamento de F\'isica, CEDENNA, Universidad de Santiago de Chile (USACH), Avenida V\'ictor Jara 3493, 9170124, Santiago, Chile.}


\author{Guillermo Romero \orcidB{}}
\email[G. Romero]{\qquad guillermo.romero@usach.cl}
\affiliation{Departamento de F\'isica, CEDENNA, Universidad de Santiago de Chile (USACH), Avenida V\'ictor Jara 3493, 9170124, Santiago, Chile.}

\author{Juan Carlos Retamal \orcidC{}}
\affiliation{Departamento de F\'isica, CEDENNA, Universidad de Santiago de Chile (USACH), Avenida V\'ictor Jara 3493, 9170124, Santiago, Chile.}

\begin{abstract}
We present a protocol for generating multiqubit quantum states with translationally invariant pairwise entanglement. Our approach is tailored for digital quantum computers with restricted qubit connectivity, a common limitation in state-of-the-art hardware platforms. We examine two configurations: star connectivity, which enables rotationally invariant entanglement, and linear connectivity, which achieves translationally invariant entanglement. For the linear configuration, we use a variant of the time-dependent density matrix renormalization group (tDMRG) algorithm to demonstrate that our protocol is independent of the qubits' number. A slight modification of the protocol reveals the presence of quantum states that exhibit periodicity of entanglement among nearest-neighbor qubits. The configurations and protocols of this work are well-suited for near-term quantum devices, offering a feasible route for the experimental realization of symmetric entangled states. 
\end{abstract}

\maketitle

\section{Introduction}
Translationally invariant quantum states are intimately connected to fundamental properties of many-body systems, such as spin chains with periodic or open boundary conditions~\cite{Dutta2015Jan,Zeng}. In such systems, the ground state preserves the translational symmetry of the Hamiltonian. Consequently,  correlations and entanglement between constituent pairs remain invariant under spatial translations. Translational symmetry also plays a crucial role in the characterization of quantum correlations, phase transitions, and topological order~\cite{Osterloh2002Apr, Osborne2002Apr, Osborne2002Sep, Vidal2003Jun, Wu2004Dec, Amico2008May, Werlang2010Aug, Werlang2011Jun,Ma2025PRXQuantum,Nussinov2009AnnPhys}. Therefore, the ability to prepare multipartite quantum states with a given symmetry is highly valuable for quantum simulations and characterization of complex many-body phenomena.

Several families of symmetric multipartite quantum states have been extensively studied in the context of quantum information. Among them, Dicke states have attracted considerable attention, with several protocols proposed theoretically and experimentally demonstrated in platforms such as trapped ions and superconducting circuits~\cite{Toyoda2011PhysRevA,Hume2009PhysRevA,Chen2023PhysRevLett,Kasture2018PhysRevA,Yu2024arXiv,Lopez2007PhysRevA}. Despite this progress, developing scalable and efficient methods for their generation remains a challenging and open problem. 

Beyond Dicke states, another important class of multipartite states is the cyclic states. Wootters~\cite{O'Connor2001Apr} was among the first to study these states in detail, analyzing entanglement in the eigenstates of symmetric Hamiltonians, such as antiferromagnetic spin rings. In their contribution, they showed that certain configurations can achieve optimal nearest-neighbor concurrence \cite{Wootters1998,Wootters2001}. He also considered the implications of translationally invariant mixed states on entanglement compared to pure states. He found values of the nearest-neighbor concurrences that are known to be attainable and that appear to be optimal for some number of particles $N$. Later, several works have analyzed this problem in different contexts~\cite{Meyer2004Jun, Benatti2005Sep, Poulsen2006May, Meill2019Oct, Hiesmayr2006Mar}. More recently, Burchardt et al.~\cite{Burchardt2021Aug} proposed a group-theoretic method to construct highly symmetric, genuinely multipartite entangled states such as Dicke-type states, presenting an example for $N = 5$ qubits implemented on an IBMQ device. The efficient preparation of symmetric entangled states, such as Dicke and cyclic states, remains an active area of research, with recent approaches utilizing auxiliary systems and measurement-based strategies~\cite{Nepomechie2024, Stojanovic2023Jul, Wang2021Sep, Wu2017Jan, Piroli2021Nov, Piroli2024Dec}.

In this work, we propose a protocol to generate multiqubit quantum states with translationally invariant pairwise entanglement, quantified by concurrence. Our approach is tailored for digital quantum computers with limited qubit connectivity, a common constraint in current platforms such as superconducting circuits. We explore two configurations: a star connectivity, enabling the generation of rotationally invariant entanglement with periodic boundary conditions (ring topology), and a linear connectivity, suited for translationally invariant states with open boundary conditions (chain topology). These two configurations are commonly found in current noisy intermediate-scale quantum (NISQ) devices, where we have nearest neighbour couplings, for example, superconducting circuits and neutral atoms. Our proposal provides a feasible route for the experimental realization of symmetric entangled states, potentially enriching applications in quantum simulation and many-body physics.

\section{\label{sec:star}Star Configuration}
\begin{figure}[t]
	\centering
	\includegraphics[width=0.9\linewidth]{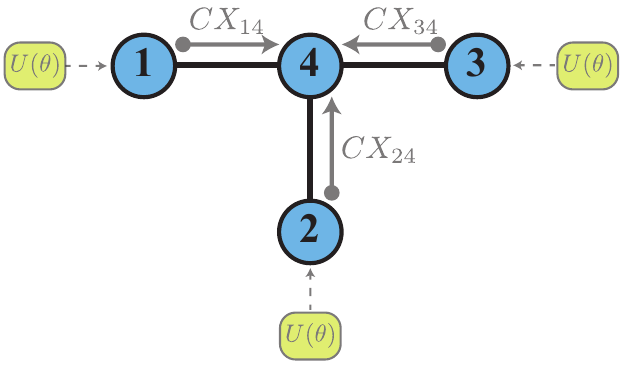}
	\caption{Star configuration of four qubits ($S_4$), illustrated as light blue circles. The protocol for generating a rotationally invariant state consists of the application of single-qubit gates $U(\theta)$, and controlled-NOT gates $CX$ using the central qubit as target and surrounding qubits as control.} 
	\label{Fig01}
\end{figure}

\begin{figure*}[t]
	\centering
	\includegraphics[width=1\linewidth]{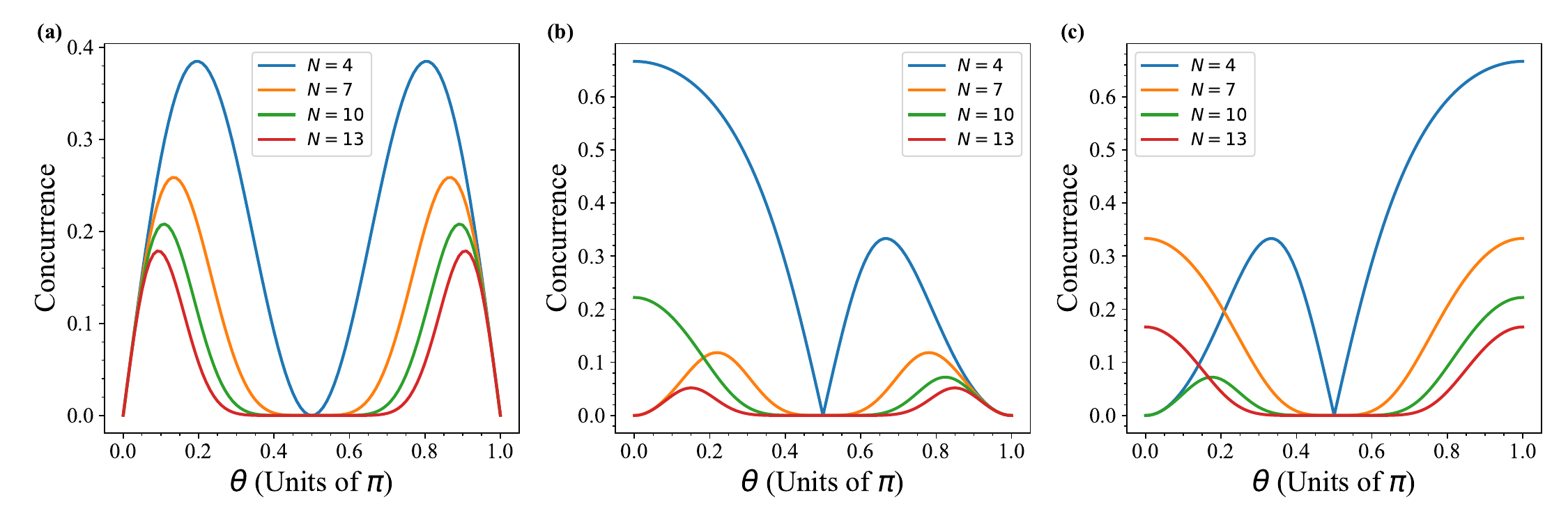}
	\caption{Concurrence in the star configuration  as a function of the single qubit gate parameter $\theta$. (a): Concurrence between the central qubit and a surronding one. (b): Concurrence between two nearest-neighbour surronding qubits after the central qubit is postselected in $\ket{0}$. (c): Concurrence between two nearest-neighbour surronding qubits after the central qubit is postselected in $\ket{1}$.} 
	\label{Fig02}
\end{figure*}

The first case of interest is the generation of rotational symmetric entangled states. To do this, we consider a central qubit connected with surrounding qubits in a star configuration ($S_N$), as depicted in Fig.~\ref{Fig01} for a four-qubit case ($S_4$). This topology can be found in various NISQ devices, including IBM's quantum chips ($S_4$), Google's Sycamore chip ($S_5$), the Zuchongzhi chip from USCT ($S_5$), and QuEra's Aquila quantum processor ($S_7$). As is described by Fig.~\ref{Fig01}, an entangling gate is applied between each external surrounding qubit and the central qubit. Each surrounding qubit is initialized in the same state $|0\rangle$ (or $|1\rangle$) and prepared in a coherent superposition via a unitary transformation:

\begin{equation}
U(\theta)=\begin{pmatrix}
a & b \\
b & -a
\end{pmatrix}
\label{unitary}
\end{equation}
where $a=\sin(\theta/2)$ and $b=\cos(\theta/2)$. This unitary operation can be decomposed in terms of native gates; for example, in the case of IBMQ devices, as $U(\theta)=\sqrt{X}R_z(\theta)\sqrt{X}$. Then, a conditional operation $CX$ is applied using the surrounding qubit as control, and the central qubit as the target. Consider a simple case of four qubits $\{q_1,q_2,q_3,q_4\}$, using  $q_4$ as the central qubit we have:
\begin{equation}
|\psi\rangle=CX_{14}\hat{U}_{1}CX_{24}\hat{U}_{2}CX_{34}\hat{U}_3 |0\rangle_1|0\rangle_2|0\rangle_3|0\rangle_4
\end{equation}
where $U_j$ is the local operations $U$ in Eq.\,(\ref{unitary}) applied to the qubit $j$, and $CX_{jk}$ denotes a controlled NOT gate where the $j$ is the control and the qubit $k$ is the target. After the protocol application, we obtain the multiqubit state:
\begin{eqnarray}
|\psi\rangle &=&a^3|0000\rangle+a^2b|0011+a^2b|0101\rangle \nonumber\\
&+&ab^2|0110\rangle+ba^2|1001\rangle+ab^2|1010\rangle \nonumber \\
&+&ab^2|1100\rangle+b^3|1111\rangle.
\end{eqnarray}
From this expression, we find that the reduced state for the qubit $q_4$ (central one) with each surrounding qubit ($q_1$, $q_2$, and $q_3$) is given by:
\begin{equation}
\rho_{4k}=\begin{pmatrix}
a^6+a^2b^4 & 0 & 0& a^5b+ab^5\\
0 & 2a^2b^ 4 & 2a^3b^3 &0\\
0 & 2a^3b^3 & 2b^2a^4 &0\\
a^5b+ab^5 & 0 & 0&a^4b^4+b^6
\end{pmatrix}.
\label{state1}
\end{equation}
Thus, we realize that every surrounding qubit is equally entangled with the central qubit. The amount of  entanglement is given by the concurrence $C(\rho)=\max\{0,\sqrt{\lambda_4}-\sqrt{\lambda_3}-\sqrt{\lambda_2}-\sqrt{\lambda_1}\}$, where $\lambda_i$ (in decreasing order) are eigenvalues of $R=\rho\tilde{\rho}$ and $\tilde{\rho}=\sigma_y\otimes \sigma_y \rho^{*}\sigma_y\otimes \sigma_y $ \cite{Wootters1998,Wootters2001}. In our case, the state $\rho_{4k}$ is given by an $X$ state \cite{Yu2006Oct,Lopez2008,Yu2009}, such that the eigenvalues of $R=\rho\tilde{\rho}$ can be easily evaluated. Writing the state $\rho_{4k}$ as:
\begin{equation}
\rho_{4k}=\begin{pmatrix}
x & 0 & 0& u\\
0 & y & \delta &0\\
0 & \delta & z &0\\
u & 0 & 0&w
\end{pmatrix},
\end{equation}
it can be shown that the eigenvalues of $\rho_{4k}$ are:
\begin{equation}
\lambda_{1\pm}=|\sqrt{xw}\pm u|\,\,\,\,\lambda_{2\pm}=|\sqrt{zy}\pm \delta| 
\end{equation}
Here, $\sqrt{xw}=ab(a^4+b^4)=u$ and $\sqrt{zy}=2a^3b^3=\delta$ so that $C(\rho)=\max\{0,2(u-\delta)\}$. It is noteworthy that the surrounding qubits are not entangled at all. 

A natural question arising at this point: Is there any procedure to equally entangle the surrounding qubits, starting from this star configuration? The answer is yes. If we consider postselection after measuring the central qubit, we get the following state for pairs $(q_1q_2)$, $(q_2,q_3)$, and $(q_1,q_3)$;

\begin{equation}
\rho_{0}=N_0\begin{pmatrix}
a^6 & 0 & 0& a^4b^2\\
0 & a^2b^4 & a^2b^4 &0\\
0 & a^2b^4 & a^2b^4 &0\\
a^4b^2 & 0 & 0&a^2b^4
\end{pmatrix}
\end{equation}

\begin{equation}
\rho_{1}=N_1\begin{pmatrix}
a^4b^2 & 0 & 0& a^2b^4\\
0 & a^4b^2 & a^4b^2 &0\\
0 & a^4b^2 & a^4b^2 &0\\
a^2b^4 & 0 & 0&b^6
\end{pmatrix}
\end{equation}
where $\rho_0 (\rho_1)$ is the state for the surrounding pairs after measuring the central qubit in state $|0\rangle$$(|1\rangle)$. Thus, we are led with a set of qubits in a ring configuration after postselecting the central qubit, embedding translational symmetry. 

We obtain similar results when considering more qubits in the star configuration, where the postselection of the central qubit provides us a rotational invariant ring state. Figure~\ref{Fig02} shows the concurrence as a function of the angle $\theta$ of the single qubit gate for the central qubit and any surrounding one (subplot a), among nearest neighbours surrounding qubits after the central qubit is postselected in $\ket{0}$ (subplot b) and after postselected in $\ket{1}$ (subplot c) considering up to 13 qubits. Also, it is essential to mention that the implementation of these states will imply considering auxiliary qubits to measure the entanglement, as has been implemented previously in Ref.~\cite{SaavedraPino2024PhysScr}. Finally, due to the limited connectivity of current NISQ processors, star configuration cannot involve a large number of qubits, implying that for large systems it is mandatory to use another more complex and scalable connectivity.

\begin{figure}[b]
	\centering
	\includegraphics[width=1.0\linewidth]{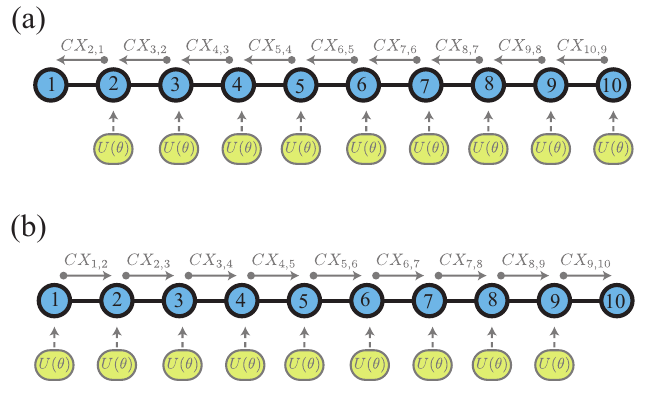}
	\caption{Linear configuration of qubits, illustrated as light blue circles. For the generation of a translational invariant many-qubit state, the sequence of single-qubit gates $U(\theta)$ and controlled-NOT gates $CX$ is applied from left to right in (a), and from right to left in (b).} 
	\label{Fig03}
\end{figure}

\section{Linear configuration}
Let us consider the generation of translationally invariant two-qubit states in a linear configuration. As is well known, one local single-qubit rotation gate and a conditional two-qubit gate are needed to entangle two qubits. On the other hand, as we learn from translationally invariant Hamiltonians, generating a state embodying the Hamiltonian symmetry could be achieved by factorizing in nearest-neighbor two-qubit gates. For a linear chain of $N$ qubits, we could initialize all the qubits in the same state, which has trivial translation-invariant symmetry. Then we could apply a local unitary operation followed by a conditional two-qubit gate sequentially as indicated in Figure~\ref{Fig03}, for a chain of $N=10$ qubits. For example, we could devise two cases from Figure \ref{Fig03}(a). Case 1: We could apply the sequence starting with a unitary $U(\theta)$ applied on qubit $q_{10}$, followed by a $CX$ from $q_{10}$ to $q_{9}$, and so on to the left. Case 2: We could start with a unitary $U(\theta)$ on qubit $q_2$, apply a $CX$ from $q_{2}$ to $q_1$, and so on to the right. Two additional possibilities are shown in Figure  \ref{Fig03}(b). Case 3: Starting with a unitary on qubit $q_1$ and a $CX$ from $q_1$ to $q_2$. Case 4: Beginning with a unitary on qubit $q_9$ and a $CX$ from  $q_9$ to $q_{10}$, and so on to the left. As shown in Appendix \ref{AppendixA}, cases 1 and 3 lead to nonentangled nearest-neighbor qubits.  

Consider case 4, as our entangling strategy, in a linear configuration of an arbitrary number of qubits with all qubits initialized in state $|0\rangle$. We sequentially apply a unitary $U(\theta)$ on qubit $q_{i}$ followed by a $CX$ gate using qubit $q_{i}$ as control and qubit $q_{i+1}$ as target, as depicted in Figure \ref{Fig03}(b), for example, a chain of $N=10$ qubits. Several questions should be answered: (a) Is the entanglement generated translationally invariant? (b) Are all neighboring qubits equally entangled? (c) How does entanglement depend on the number of qubits? (d) How can we extend to the case of an arbitrarily large number of qubits? 
\begin{figure}[t]
	\centering
	\includegraphics[width=1\linewidth]{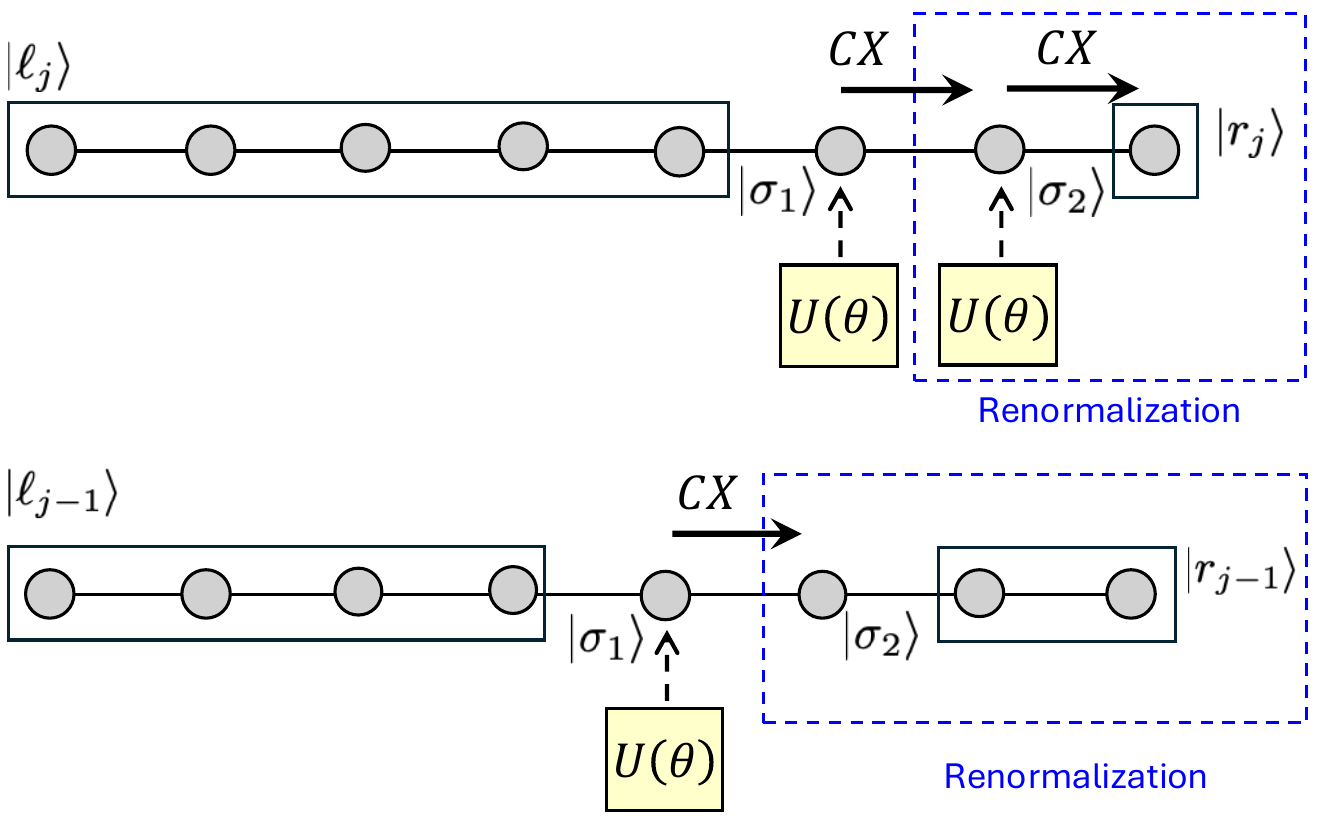}
	\caption{A schematic of the DMRG algorithm for a lattice of $N=8$ qubits. Each block containing $s$ sites and $m$ states kept is described by $\mathcal{B}(s,m)$. A local unitary applied to each qubit is represented by $U(\theta)$, whereas the two-qubit entangling gate is represented by $CX$.}  
	\label{Fig04}
\end{figure}

To answer the posed questions, let us consider a finite chain of qubits to implement one of the above-mentioned entangling strategies, and see how the entanglement behaves as we vary the number of qubits. To accomplish this goal, we use the density-matrix renormalization group (DMRG) algorithm \cite{White1992Nov,Schollwock2005Apr,Verstraete2023May} to compress the representation of the many-qubit wave function. This strategy can be viewed as a variant of the time-dependent DMRG algorithm \cite{White2004Aug,Daley2004Apr,Vidal2004Jul,Gobert2005Mar,Feiguin2011Dec} applied to quantum circuits composed of unitaries, and is useful for the approximate simulation of quantum circuits \cite{Zhou2020Nov,Ayral2023Apr}. Our DMRG algorithm can be efficiently scaled up to many-qubit systems, as its accuracy in approximating the multiqubit wave function relies on low entanglement, a condition naturally satisfied in our multiqubit system.
\begin{figure}[t]
	\centering
	\includegraphics[width=1.\linewidth]{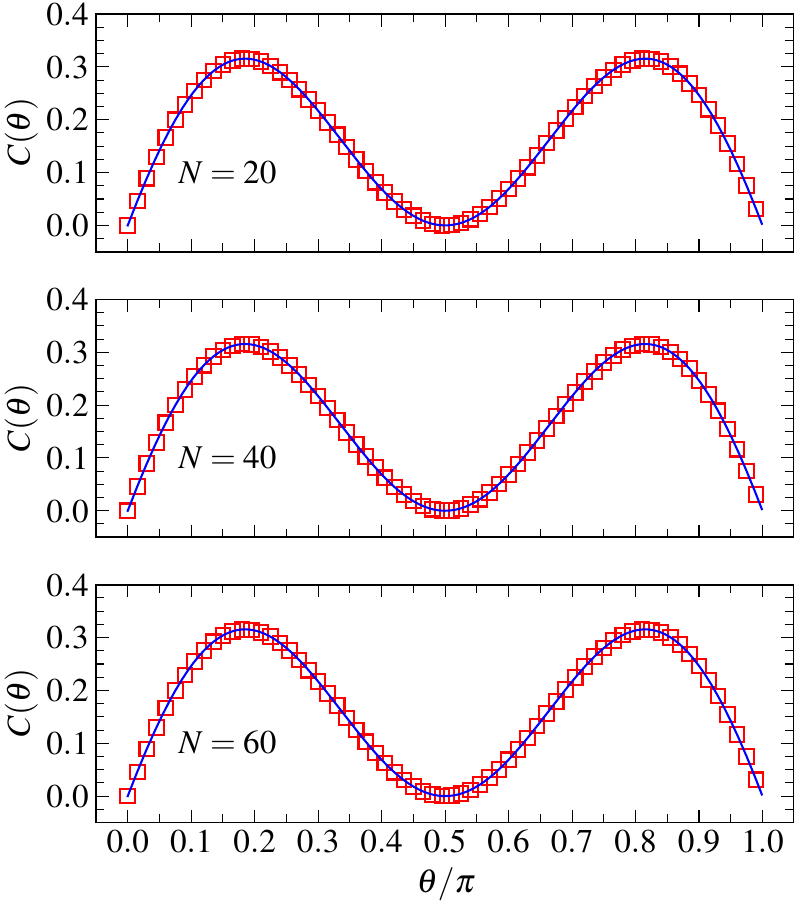}
	\caption{Pair concurrences as a function of the angle $\theta$ for a linear chain of $N=20$, $N=40$, and $N=60$ qubits. The blue curves represent the concurrence of the pair $(q_{N/2},q_{N/2+1})$, whereas red squares represent the concurrence of the pair $(q_{N-2},q_{N-1})$. The numerical simulations follow the DMRG strategy described in the main text.} 
	\label{Fig05}
\end{figure}

In our DMRG algorithm, the single-qubit and two-qubit gate sequences are applied to each pair of qubits in a single left-to-right sweep; see Fig.~\ref{Fig04} for a schematic of the DMRG algorithm considering, for simplicity, $N=8$ qubits. Here, each block, containing $s$ sites and $m$ states, is described by $\mathcal{B}(s,m)$. In the first step, the multiqubit wave function has the structure $\mathcal{B}(5,m) \bullet \, \bullet\, \mathcal{B}(1,2)$, and we apply the set of unitaries $U(\theta)$ and $CX$ as shown in the upper panel of Fig.~\ref{Fig04}. Then, we proceed with White's wave function prediction~\cite{White1996Oct}. The latter corresponds to transforming the wave function into the basis appropriate for the next DMRG algorithm iteration, and it works as follows. Let us consider the wave function ansatz at the $j$th step of the finite-system DMRG algorithm, see the upper panel of Fig.~\ref{Fig04}, 
\begin{equation}
    \ket{\psi}=\sum_{\ell_{j},\sigma_{1},\sigma_{2},r_{j}}\psi(\ell_{j}\sigma_{1}\sigma_{2}r_{j})\ket{\ell_{j}\sigma_{1}\sigma_{2}r_{j}},    
\end{equation}
where the indexes take values $\ell_{j},r_{j}=1,\hdots, m$, and $\sigma_1,\sigma_2=0,1$. Now, we need the wave function $\ket{\psi}$ in the basis state $\ket{\ell_{j-1}\sigma_1\sigma_2r_{j-1}}$ to apply the local unitary $U(\theta)$ and $CX$; see the lower panel of Fig.~\ref{Fig04}. The basis change is obtained as
\begin{eqnarray}
    \ket{\psi}&=\sum_{\ell_{j-1}\sigma_1\sigma_2r_{j-1}}\psi(\ell_{j-1}\sigma_{1}\sigma_{2}r_{j-1})\ket{\ell_{j-1}\sigma_1\sigma_2r_{j-1}},\nonumber\\
    &\psi(\ell_{j-1}\sigma_{1}\sigma_{2}r_{j-1})=\sum_{\ell_{j}\sigma'_2r_j}\psi_{\ell_{j}\sigma_{2}\sigma'_{2}r_{j}}(O_L)_{\ell_{j-1}\sigma_1,\ell_j}(O_R)^{\dag}_{r_{j-1},\sigma_2'r_j}\nonumber\\
\end{eqnarray}
where $(O_L)_{\ell_{j-1}\sigma_1,\ell_j}=\braket{\ell_{j-1}\sigma_1}{\ell_j}$ and $(O_R)_{\sigma_2r_j,r_{j-1}}=\braket{\sigma_2r_j}{r_{j-1}}$ are the left and right isometric matrices, respectively. Each isometry contains in its columns the eigenvectors of the left and right reduced density matrices obtained from renormalization, in descending order according to their eigenvalues $\lambda_1>\lambda_2>\hdots>\lambda_m$. For each angle $\theta$ in Eq.~(\ref{unitary}), we iterate the DMRG algorithm until reaching the left-most site of the qubit array.    

\begin{figure}[t]
	\centering
	\includegraphics[width=1\linewidth]{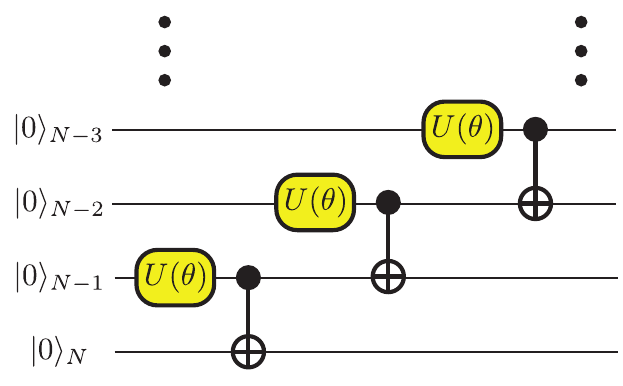}
	\caption{Quantum circuit for linear configuration with backward protocol, c.f. Fig.~\ref{Fig03}(b).} 
	\label{Fig06}
\end{figure}

The first main result for a chain of $N$ qubits with open boundary conditions is depicted in Fig.~\ref{Fig05}, where we plot the concurrence between qubit pairs in the central bond (blue curve) and qubit pairs in the penultimate bond (red squares), for a lattice of $N=20$, $N=40$, and $N=60$ qubits. Here, the quantum circuit for translationally invariant entanglement generation is depicted in Fig.~\ref{Fig06}. This quantum circuit builds upon a single layer of two-qubit gates. From Fig.~\ref{Fig05} we see that the amount of entanglement is translationally invariant for the bulk qubits and is independent of the number of qubits. 
Once we learn from numerical simulations under the DMRG strategy that the bulk entanglement is independent of the number of qubits, we can analytically evaluate the entanglement in the simplest problem of $N=6$ qubits that exhibit three internal bulk pairs. In such a case, we consider the following protocol:

\begin{equation}
|\psi\rangle=CX_{12}\hat{U}_{1}...CX_{34}\hat{U}_{3}CX_{45}\hat{U}_{4}CX_{56}\hat{U}_5 |0\rangle_1|0\rangle_2|0\rangle_3|0\rangle_4|0\rangle_5|0\rangle_6
\end{equation}
Considering the local unitary given by Eq.\,(\ref{unitary}), we realize that the reduced quantum state for bulk qubit pairs $(q_2,q_3),(q_3,q_4),(q_4,q_5)$ is given by
\begin{equation}
\rho_{}=\begin{pmatrix}
x& 0 & 0& r\\
0 & w& z&0\\
0 &z & w &0\\
r& 0 & 0&w
\end{pmatrix}
\label{state1}
\end{equation}

\begin{eqnarray}
x&=&a^6+b^6  \nonumber\\
w&=&a^2b^2\nonumber \\
r&=& a^5b+ab^5 \nonumber \\
z&=&2(a^5b^3+a^3b^5)
\end{eqnarray}
where we use the condition $a^2+b^2=1$. A more involved calculation can be carried out for 8 qubits, containing five bulk pairs, leading to the same result.  The Wootters concurrence for bulk pairs can be calculated from the state in Eq.\,(\ref{state1}), resulting in the expression $C(\theta)=\max\{0,C_{\pm}(\theta\}$, where:

\begin{equation}
C_{\pm}(\theta) = \frac{1}{8}\bigg[-2 + 2\cos{(2\theta)}\pm(5\sin{(\theta)}+\sin{(3\theta))}\bigg]
\label{ConcL}
\end{equation}
Additionally, it can be shown that for external extreme pairs, the entanglement is given by:
\begin{equation}
C(\theta) = |\sin(\theta)\cos(\theta)|
\label{Concextreme}
\end{equation}
\begin{figure}[t]
	\centering
	\includegraphics[width=1.\linewidth]{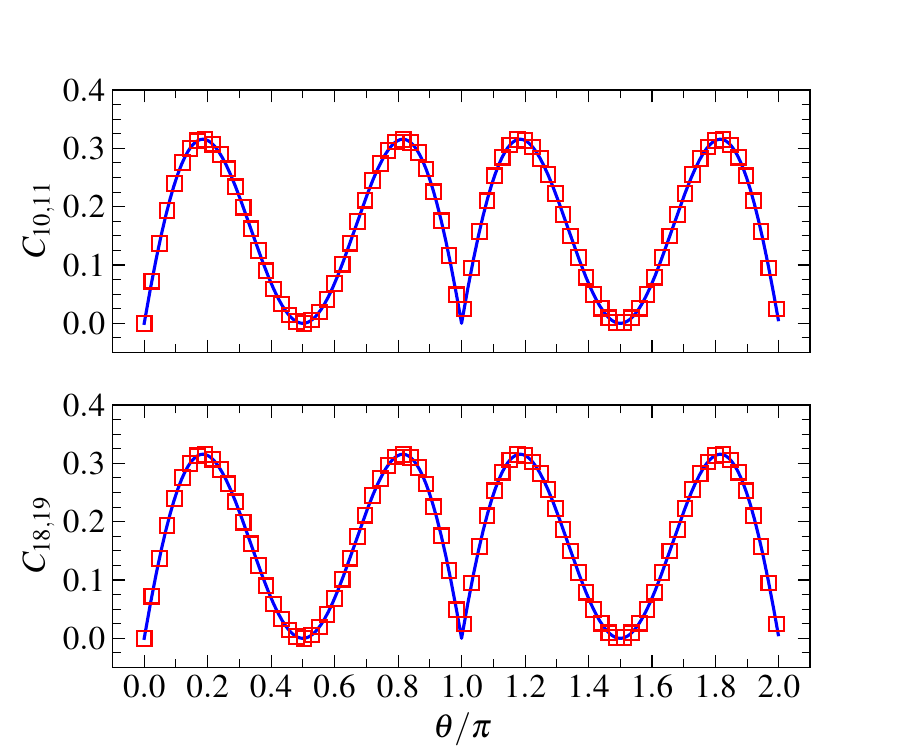}
	\caption{Pair concurrences as a function of the angle $\theta$ for a linear chain of $N=20$ qubits. The blue curves represent the concurrence numerically computed using the DMRG algorithm, whereas red squares represent the analytical concurrence of Eq.\,(\ref{ConcL}). The upper panel shows the concurrence of the pair $(q_{10},q_{11})$, and the lower panel shows the concurrence of the pair $(q_{18},q_{19})$.} 
	\label{Fig07}
\end{figure}
Figure \ref{Fig07} shows the concurrence for $\theta$ in the interval $[0,2\pi]$,  calculated for different pairs for a finite chain of $N=20$ qubits following the quantum circuit of Fig.~\ref{Fig06}. The blue curves represent the numerical simulation using DMRG. In contrast, the red squares represent the analytical pair concurrence given by Eq.\,(\ref{ConcL}). The analytical formula is in complete agreement with the numerical DMRG result. The concurrence is invariant in the bulk of the qubit chain; however, the concurrences between the extreme pairs $(q_{1},q_{2})$ and $(q_{19},q_{20})$ are equal but differ from those of the internal pairs. As can be directly inferred from Eq.\,(\ref{ConcL}) minimal values for $C$ occur for $\theta = 0,\pi/2,\pi,3\pi/2,2\pi$. The maximal entanglement for a linear chain in this case occurs for $\sin \theta=(-1+\sqrt{7})/3$, which corresponds to $\theta=0.1848\pi$. The remaining peaks are located in $\theta=\pi\pm0.1848\pi,2\pi-0.1848\pi$.
\begin{figure}[t]
	\centering
	\includegraphics[width=1.\linewidth]{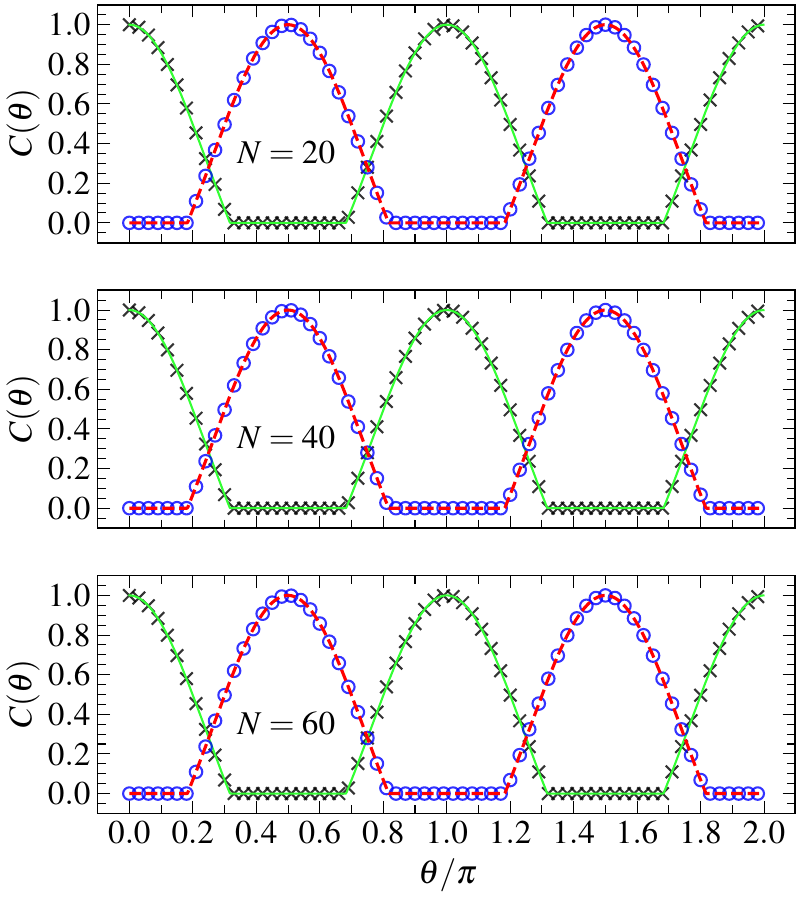}
	\caption{Periodic configuration. In each panel, the continuous green line corresponds to the concurrence for pairs $(q_{2},q_{3}),(q_{4},q_{5}),... $ etc., and crosses stand for the analytical expression of Eq.\,(\ref{Concper1}). The dashed red line corresponds to the concurrence for pairs $(q_{3},q_{4}),(q_{5},q_{6}),... $ etc, whereas circles stand for the analytical expression of Eq.\,(\ref{Concper2}).} 
	\label{Fig08}
\end{figure}
\section{Periodic linear configuration}
An interesting problem would be to find quantum states that exhibit periodicity of entanglement among nearest-neighbor qubits. A strategy for implementing these types of states assumes that instead of using the same unitary, we could consider that two unitary operations $\hat{U}_a$ and $\hat{U}_b$ are applied on the qubit chain alternating them, as suggested in the following expression for a chain of $n$-qubits:
\begin{equation}
|\psi\rangle=CX_{12}\hat{U_a}_{1}...CX_{n-2n-1}\hat{U_b}_{n-2}CX_{nn-1}\hat{U_a}_{n-1} |0\rangle_1..|0\rangle_{n-1}|0\rangle_n
\end{equation}
Analytical calculations are not difficult to implement for a sequence of  $N=8$ qubits, where unitary operations $U_{ak}$ and $U_{bk}$ are given by Eq.\,(\ref{unitary}) with $a_{1(2)}\,=\,\sin(\theta_{1(2)}/2)$ and $b_{1(2)}\,=\,\cos(\theta_{1(2)}/2)$. We find that for qubits pairs $(q_{2},q_{3}),(q_{4},q_{5}),(q_{6},q_{7})$ the state is given by:
\begin{equation}
\rho_{}=\begin{pmatrix}
x& 0 & 0& r\\
0 & y& z&0\\
0 &z & y &0\\
r& 0 & 0&w
\end{pmatrix}
\end{equation}
where we defined 
\begin{eqnarray}
x&=&a{_1}^4a_{2}^2+b{_1}^4b_{2}^2  \nonumber\\
y&=&a_{1}^2b_{1}^2\nonumber \\
w&=&a{_1}^4b_{2}^2+b{_1}^4a_{2}^2 \nonumber \\
r&=& a_{2}b_{2}(a_{1}^4+b_{1}^4) \nonumber \\
z&=&2a_{1}^2a_{2}b_{1}^2b_{2}),
\end{eqnarray}
whereas for pairs $(q_{3},q_{4}),(q_{5},q_{6})$ the state is given by:
\begin{equation}
\rho_{}=\begin{pmatrix}
x'& 0 & 0& r'\\
0 & y'& z'&0\\
0 &z' & y' &0\\
r'& 0 & 0&w'
\end{pmatrix}
\end{equation}
where we defined 
\begin{eqnarray}
x'&=&a{_1}^2a_{4}^2+b{_1}^2b_{2}^4  \nonumber\\
y'&=&a_{2}^2b_{2}^2\nonumber \\
w'&=&a{_1}^2b_{2}^4+b{_1}^2a_{4}^2 \nonumber \\
r'&=& a_{1}b_{1}(a_{2}^4+b_{2}^4) \nonumber \\
z'&=&2a_{1}a_{2}^2b_{1}b_{2}^2
\end{eqnarray}
Thus, for a chain of $N=8$ qubits, the bulk qubit states exhibit periodicity. It is hard to calculate these analytical results when increasing $N$.  However, the numerical analysis for a chain of $N=20,40,60$ qubits is shown in Fig.~\ref{Fig08}, where the entanglement exhibits periodicity for consecutive pairs of bulk nearest-neighbor qubits. This result is independent of the number of qubits. It can be shown that the entanglement for pairs $(q_2,q_3)$ and $(q_4,q_5)$ is given by the expression $C_{23}= \max\{0,C_{23\pm}\}$ :
\begin{equation}
C_{23\pm}=\frac{1}{4}(-1+\cos2\theta_2\pm(3+\cos{2\theta_2})\sin\theta_1)
\label{Concper1}
\end{equation}
 and for the pair $(q_3,q_4)$ is given by $C_{34}= \max\{0,C_{34\pm}\}$ where
 \begin{equation}
C_{34\pm}=-2(\sin\theta_1/2\cos\theta_1/2)^2\pm \frac{1}{4}(3+\cos2\theta_1)\sin\theta_2
\label{Concper2}
\end{equation}
These expressions are reduced to Eq.\,(\ref{ConcL}) for $\theta_1=\theta_2$ as expected. 
Figure \ref{Fig08} shows the entanglement of bulk qubit pairs for different extensions of the qubit chain. The continuous green line corresponds to the concurrence for pairs $(q_{2},q_{3}),(q_{4},q_{5}),..$ etc., and the dashed red line to pairs $(q_{3},q_{4}),(q_{5},q_{6}),..$ etc.  As in the previous case, the entanglement is independent of the chain extension. In this extreme case, for $\theta_2=\theta_1+\pi/2$, we observe that consecutive pairs are dimerized. As one pair is completely entangled, the neighboring pair is completely disentangled. This disentanglement in terms of angle $\theta$ resembles the feature of sudden death of entanglement and sudden birth of entanglement \cite{Lopez2008}, as they appear as an abrupt change of entanglement. 
\section{Conclusion}
In this work, we addressed the problem of generating multipartite entangled quantum states that exhibit translational invariance. This means that the quantum state and, consequently, the entanglement between neighboring pairs of qubits is the same. We explored star configurations where a central qubit is entangled with qubits surrounding it, such that each outer qubit has the same entanglement with the central one. From this configuration, through a measurement process on the central qubit, ring configurations can be obtained where all pairs of qubits become equally entangled. Additionally, we investigated a linear configuration, where we discovered that, under suitable entangling gates, entangled states with translational invariance can be achieved. In this case, the entanglement does not depend on the number of qubits. Additional features of translationally invariant entangled states have been explored, such as considering the application of entangling gates in a periodic fashion, thus obtaining a periodic structure for the states and the entanglement between adjacent pairs. Using a protocol tailored for digital quantum computers with restricted qubit connectivity, our results offer a feasible route for the experimental realization of symmetric entangled states in state-of-the-art quantum platforms.

\section*{Acknowledgments}
We acknowledge financial support from Agencia Nacional de Investigaci\'on y Desarrollo (ANID): Centro de Investigación Asociativa ANID CIA 250002, Fondecyt grant 1231172; and  Universidad de Santiago de Chile: DICYT Asociativo Grant No. 042431AA$\_$DAS.

\bibliography{Mybib}

\newpage
\appendix
\section{Non entangled qubit states}
\label{AppendixA}
We consider case 1, see Fig.~\ref{Fig03}(a), and apply the sequence starting with a unitary $U(\theta)$ applied on qubit $q_{10}$, followed by a $CX$ from $q_{10}$ to $q_{9}$, and so on to the left. The result is independent of the number of qubits, so we outline the result for three qubits. We call the final state $\ket{\Psi_n}$, where the subindex refers to the number of qubits involved. For two and three qubits, we have 
\begin{eqnarray}
    \ket{\Psi_2}&=&c\ket{00}+s\ket{11},\nonumber\\
    \ket{\Psi_3}&=&c^2\ket{000}+cs(\ket{011}+\ket{111})-s^2\ket{100}\nonumber\\
    &=&\underbrace{(c^2\ket{0}-s^2\ket{1})}_{\ket{\phi_1^{00}}}\ket{00}+\underbrace{cs(\ket{0}+\ket{1})}_{\ket{\phi_0^{11}}}\ket{11}
\end{eqnarray}
where $c=\cos(\theta/2)$ and $s=\sin(\theta/2)$. We note that the states $\ket{\phi_1^{00}}$ and $\ket{\phi_1^{11}}$ are not normalized and not orthogonal. If we trace the last qubit, the reduced density matrix of the first and the second one is:
\begin{eqnarray}
    \rho_{1,2}
    =&&\ketbra{\phi_1^{00}}{\phi_1^{00}}\otimes\ketbra{0}{0}+\ketbra{\phi_1^{11}}{\phi_1^{11}}\otimes\ketbra{1}{1}
\end{eqnarray}
obtaining a separable matrix; therefore, the first and the second qubits do not have entanglement. Now, if we trace the first one, we obtain the next density matrix for the last two:
\begin{eqnarray}
    \rho_{2,3}=&&(c^4+s^4)\ketbra{00}{00} + 2c^2s^2\ketbra{11}{11} \nonumber\\
    &&+ cs(c^2-s^2)(\ketbra{00}{11}+\ketbra{11}{00}).
\end{eqnarray}
Considering $\alpha=c^4+s^4$, $\beta=cs(c^2-s^2)$, and $\gamma=2c^2s^2$, the reduced density matrix can be written as the next effective $2\times 2$ matrix:
\begin{eqnarray}
    \rho_{2,3}=\begin{pmatrix}
        \alpha & \beta\\
        \beta & \gamma
    \end{pmatrix}
\end{eqnarray}
then, the concurrence among the last two qubits takes the form
\begin{eqnarray}
    \mathcal{C}=2|\beta|=2|cs(c^2-s^2)|.
\end{eqnarray}

If we consider four qubits, the state after the protocol is given by
\begin{eqnarray}
    \ket{\Psi_4}=&&\underbrace{\left(c\ket{\phi_1^{00}}\ket{0}-s\ket{\phi_1^{11}}\ket{1}\right)}_{\ket{\phi_2^{00}}}\ket{00}\nonumber\\
    &&+\underbrace{\left(s\ket{\phi_1^{00}}\ket{0}+c\ket{\phi_1^{11}}\ket{1}\right)}_{\ket{\phi_2^{11}}}\ket{11}
\end{eqnarray}
obtaining the same structure as in the case of three qubits. In general, we can show that
\begin{eqnarray}
    \ket{\Psi_n}=&&\ket{\phi_{n-2}^{00}}\ket{00}+\ket{\phi_{n-2}^{11}}\ket{11}
    \label{GeneralN}
\end{eqnarray}
with
\begin{eqnarray}
    \ket{\phi_{n}^{00}}=&&c\ket{\phi_{n-1}^{00}}\ket{0}-s\ket{\phi_{n-1}^{11}}\ket{1},\nonumber\\
    \ket{\phi_{n}^{11}}=&&s\ket{\phi_{n-1}^{00}}\ket{0}+c\ket{\phi_{n-1}^{11}}\ket{1}.
\end{eqnarray}

We observe that if we trace the last qubit from Eq.\,(\ref{GeneralN}), we find that the $(n-1)$th qubit is disentangled from the rest; specifically, qubits $n-1$ and $n-2$ are not entangled. Now, if we trace the last two qubits from Eq.\,(\ref{GeneralN}), we get a density matrix given by:
\begin{eqnarray}
\rho=&&\ketbra{\phi_{n-2}^{00}}{\phi_{n-2}^{00}}+\ketbra{\phi_{n-2}^{11}}{\phi_{n-2}^{11}}\nonumber\\
    =&&\ketbra{\phi_{n-3}^{00}}{\phi_{n-3}^{00}}\otimes\ketbra{0}{0}+\ketbra{\phi_{n-3}^{11}}{\phi_{n-3}^{11}}\otimes\ketbra{1}{1},
    \label{n-3}
\end{eqnarray}
where the $(n-2)$th qubit is disentangled from the rest. Specifically, qubits $n-2$ and $n-3$ are not entangled. For Eq.\,(\ref{n-3}), we can infer that always qubits $j$ and $j+1$ are not entangled except for $n-1$ and $n$, namely, the end of the chain.

\end{document}